\documentclass[preprint,12pt]{elsarticle}
\usepackage{amssymb}

\journal{J. Parallel Distrib. Comput.}

\begin{document}

\begin{frontmatter}



\title{Python Non-Uniform Fast Fourier Transform (PyNUFFT): multi-dimensional non-Cartesian image reconstruction package for heterogeneous platforms and applications to MRI}

\author[label1,label2]{Jyh-Miin Lin}
\address[label1]{Department of Radiology, University of Cambridge, Cambridge CB2 0QQ, UK jml86@cam.ac.uk}
\address[label2]{Graduate Institute of Biomedical Electronics and Bioinformatics, National Taiwan University, Taipei, Taiwan}

\begin{abstract}
This paper reports the development of a Python Non-Uniform Fast Fourier Transform (PyNUFFT) package, which accelerates non-Cartesian image reconstruction on heterogeneous platforms. Scientific computing with Python encompasses a mature and integrated environment. The NUFFT algorithm has been extensively used for non-Cartesian image reconstruction but previously there was no native Python NUFFT library. The current PyNUFFT software enables multi-dimensional NUFFT on heterogeneous platforms. The PyNUFFT also provides several solvers, including the conjugate gradient method, $\ell$1 total-variation regularized ordinary least square (L1TV-OLS) and $\ell$1 total-variation regularized least absolute deviation (L1TV-LAD). Metaprogramming libraries were employed to accelerate PyNUFFT. The PyNUFFT package has been tested on multi-core CPU and GPU, with acceleration factors of 6.3 - 9.5$\times$ on a 32 thread CPU platform and 5.4 - 13$\times$ on the GPU.

\end{abstract}


\end{frontmatter}


\clearpage
{\bf Highlights}
\begin{itemize}
\item A Python non-uniform fast Fourier transform (PyNUFFT) package was developed.
\item Computations are accelerated on heterogeneous systems, including multi-core CPU and GPU.

\item It provides the accelerated nonlinear conjugate gradient method, $\ell$1 total-variation regularized ordinary least square and $\ell$1 total-variation regularized least absolute deviation.

\item Pre-indexing enables multi-dimensional NUFFT and fast image gradients on heterogeneous platforms. 

\item The single-precision version is dual-licensed under MIT and LGPL-3.0. 

\item Applications to magnetic resonance imaging reconstruction are presented.

\end{itemize}
\clearpage
 
\section{Introduction}
   Fast Fourier transform (FFT) is an exact fast algorithm to compute discrete Fourier transform when data are acquired on an equispaced grid. In certain image processing fields, however, the frequency locations are irregularly distributed, which obstructs the use of FFT. The alternative Non-Uniform Fast Fourier Transform (NUFFT) algorithm is a fast mapping for computing non-equispaced frequency components. Several previous non-Cartesian image reconstructions are summarized in the discussion section (Section 4.1).
     
Python is a fully-fledged and well-supported programming language in data science. The importance of Python language is manifested in the recent surge of interest in machine learning. Developers have increasingly relied on Python to build software, taking advantage of its abundant libraries and active community. Yet the standard Python numerical environments lack a native implementation of NUFFT packages, and the development of an efficient Python NUFFT may fill a gap in the image processing field. However, Python has been notorious for its slow execution speed, which hampers the implementation of an efficient Python NUFFT.

During the past decade, the speed of Python has been greatly improved by numerical libraries with rapid array manipulations and vendor-provided performance libraries. However, parallel computing using multi-threading model cannot be easily implemented in Python. This problem is mostly due to the Global Interpreter Lock (GIL) of Python interpreter, which only allowed one core to be used at the same time, while the multi-threading capabilities of modern Symmetric multiprocessing (SMP) processors cannot be exploited.

Recently, general purpose graphic processing unit (GPGPU) computing has enabled enormous acceleration by offloading computations to hundreds to several thousands of parallel processing units on GPUs. This emerging programming models may enable one to develop an efficient Python NUFFT package, by circumventing the limitations of GIL. Two GPGPU architectures are commonly used, i.e. the proprietary Compute Unified Device Architecture (CUDA, NVIDIA, Santa Clara, CA, USA) and Open Computing Language (OpenCL, Khronos Group, Beaverton, OR,USA). These two similar architectures can be ported to each other, or could be dynamically generated from Python codes \cite{klockner2012pycuda}.

Thus, we have developed the Python NUFFT (PyNUFFT) package, which was optimized for heterogeneous systems, including multi-core CPU and GPU. The design of PyNUFFT aims to reduce the run-time while maintaining the readability of Python program. Python NUFFT (PyNUFFT) contains the following features: (1) algorithms written and tested for heterogeneous platforms (including multi-core CPU and GPU); (2) pre-indexing to handle multi-dimensional NUFFT and image gradient; (3) providing several nonlinear solvers.

\section{Material and methods}

 PyNUFFT software was developed on a 64-bit Linux system with a quad-core CPU (Intel i7-6700HQ Processor, 6M Cache, up to 3.50 GHz) and a GPU (NVIDIA Geforce GTX 965m, 1024 CUDA Cores at 944+ BoostBase Clock(MHz), 2GB GDDR5 memory at 2500 MHz Memory Clock). The current version was modified from the previous version which was accelerated with CUDA \cite{lin2017bbmc_pynufft}. The new version is built on PyOpenCL, PyCUDA \cite{klockner2012pycuda} and Reikna  \cite{reikna} licensed under dual MIT and GNU Lesser General Public License v3.0 (LGPL-3.0) \cite{gplv3}. Thus, PyNUFFT may be used in a variety of projects.

  \subsection{PyNUFFT: an NUFFT implementation in Python}
\begin{figure}[h]
\centering
\includegraphics[width=1.2\linewidth]{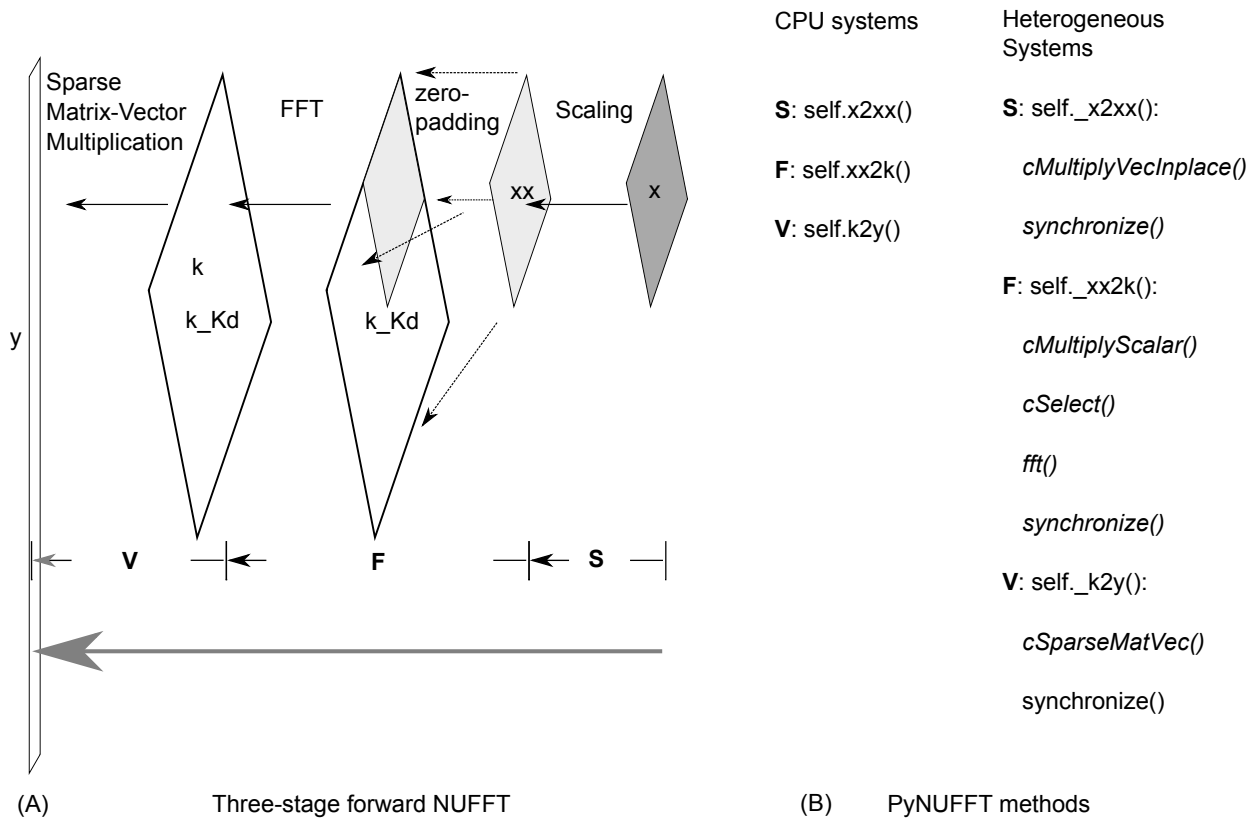} \caption{(A) A 2D example of the forward PyNUFFT (B) The methods provided in PyNUFFT. Forward NUFFT can be decomposed into three stages: scaling, FFT, and interpolation.}
\label{fig:fig1}
\end{figure}

  The execution of PyNUFFT proceeds through three major stages: (1) scaling, (2) oversampled FFT, and (3) interpolation. The three stages can be formulated as the combination of linear operations:
     \begin{eqnarray}
     \bf A = V F S
     \end{eqnarray} 
     where $\bf A$ is the NUFFT, $\bf S$ is the scaling, $\bf F$ is the FFT, and $\bf V$ is the interpolation. 
     The default kernel function in PyNUFFT is the min-max kernel \cite{fessler2003nonuniform}, whereas other kernel functions can be used. 
     The design of the three-stage NUFFT algorithm is illustrated in Figure \ref{fig:fig1}(A). Batch-mode NUFFT can be performed with serial processing or parallel processing.
     
\subsubsection{Scaling}

	Scaling was performed by in-place multiplication cMultiplyVecInplace. The complex multi-dimensional array was offloaded to the device. 
	
	\subsubsection{Oversampled FFT}
	This stage is composed of two steps: (1) zero-padding which copies the small array to a larger array, and (2) FFT. The first step recomputed the array indexes on-the-flight with logical operations, a step  which is not well supported on GPUs. 
	Here, a generalized pre-indexing procedure was implemented to avoid matrix reshaping on the flight, and the cSelect subroutine copies the array1 to array2 according to the pre-indexes order1 and order2. This pre-indexing avoids multi-dimensional matrix reshaping on the flight, thus greatly simplifying the algorithm on GPU platforms. In addition, pre-indexing can be generalized to multi-dimensional arrays (with size $N_{in}$, $N_{out}$).

 Once the indexes (inlist, outlist) were obtained, the input array can be rapidly copied to a larger array. No matrix reshaping will be needed during iterations. 
 
 The inverse FFT is also based on the same subroutines of the oversampled FFT, but it reverses the order of computations. Thus, an IFFT is followed by array copying.
 
 	\subsubsection{Interpolation}		
	While the current PyNUFFT includes the min-max interpolator \cite{fessler2003nonuniform}, other kernels could also be used. The scaling factor of the min-max interpolator was designed to minimize the error of off-grid samples \cite{fessler2003nonuniform}. The interpolation kernel was stored as the Compressed Sparse Row matrix (CSR) format for C-order (row-major) array. Thus, the indexing was accommodated for C-order and the cSparseMatVec subroutine could quickly compute the interpolation without matrix reshaping. cSparseMatVec had been optimized to exploit the data coalescence and parallel threads on the heterogeneous platforms. The warp in CUDA (or wavefront in OpenCL) control the size of workgroups in the cSparseMatVec kernel. Note that the indexing of the C-ordered array is different from the fortran array (F-order) implemented in MATLAB.
	
	Gridding is the conjugate transpose of the interpolation, which also uses the same cSparseMatVec subroutine.
   \subsubsection{Adjoint PyNUFFT}
\begin{figure}[!h]
\centering
\includegraphics[width=1.2\linewidth]{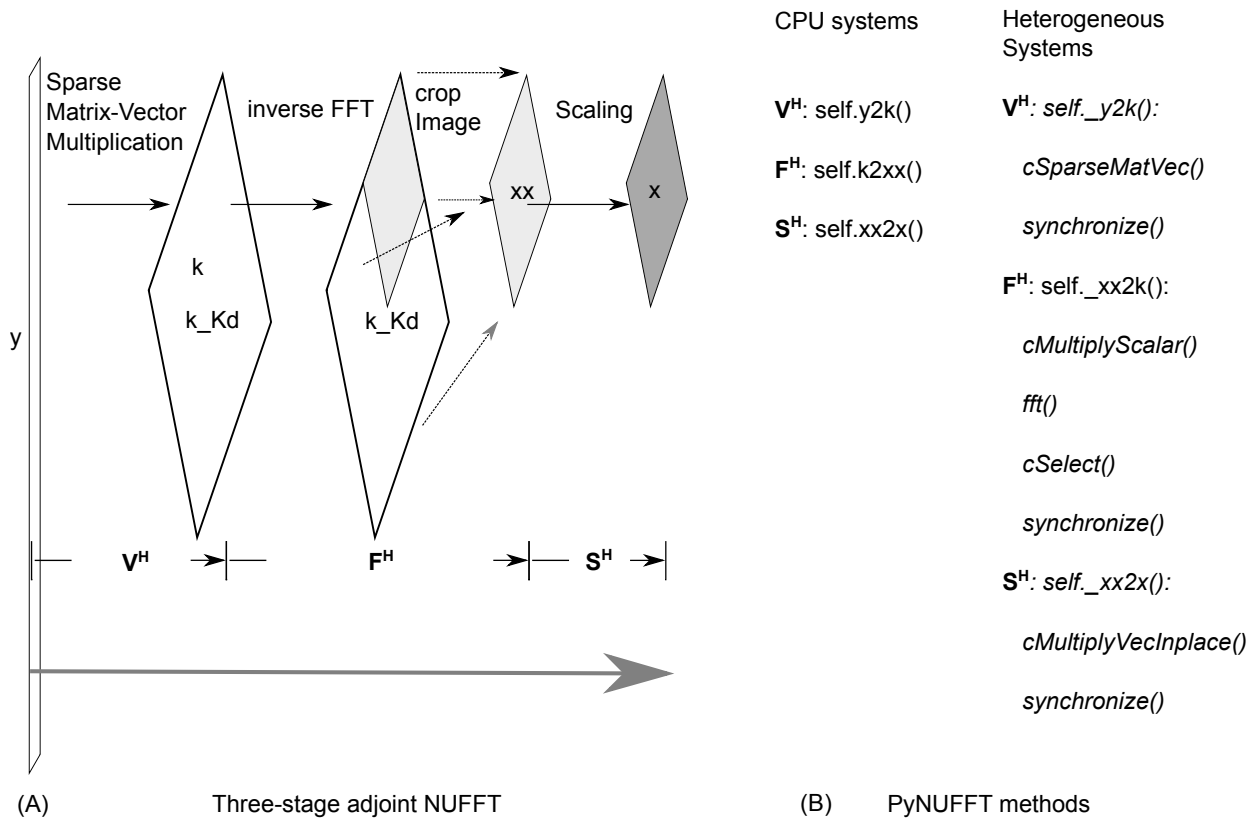} \caption{(A) The adjoint PyNUFFT (B) The methods provided in PyNUFFT. Adjoint NUFFT can be decomposed into three stages: gridding, IFFT, and rescaling.}
\label{fig:adjoint}
\end{figure}

Adjoint NUFFT reverses the order of the forward NUFFT. Each stage is the conjugate transpose (Hermitian transpose) of the forward NUFFT:
     \begin{eqnarray}	
     \bf A^H = S^HF^H V^H
     \end{eqnarray}
which is also illustrated in Figure \ref{fig:adjoint}.

\subsubsection{Self-adjoint NUFFT (Toeplitz)} 
In iterative algorithms, the cost function is used to represent data consistency:
\begin{eqnarray}
cost\ function: argmin_x \left\Vert {\bf A}x - y \right\Vert_2\\
\end{eqnarray}
in which the normal equation is composed of interpolation ($\bf A$) and gridding ($\bf A^H$):
\begin{eqnarray}
normal\ equation: {\bf A^HA}x = {\bf A^H}y 
\label{eq:normalequation}
\end{eqnarray}

\begin{figure}[h]
\centering
\includegraphics[width=1.1\linewidth]{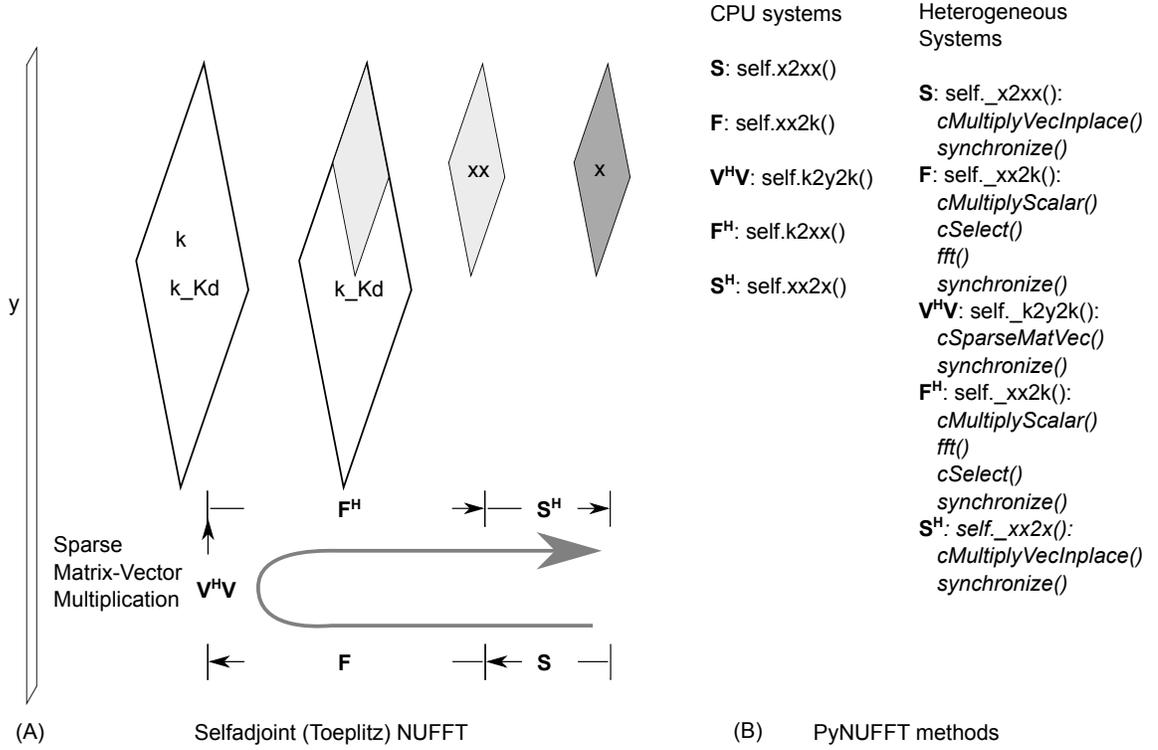} \caption{(A) The self-adjoint PyNUFFT (Toeplitz) (B) The methods are provided in PyNUFFT. The Toeplitz method precomputes the interpolation-gridding matrix ($\bf V^HV$), which can be accelerated on CPU and GPU. See Table \ref{table:accelerated_transforms} for measured acceleration factors.}
\label{fig:selfadjoint}
\end{figure}

Thus, precomputing the ${\bf A^HA}$ can improve the run-time efficiency\cite{fessler2005toeplitz}. 
See Figure \ref{fig:selfadjoint}

\subsection{Solver}
 	The NUFFT provided solvers to restore multi-dimensional images (or 1D signals in the time-domain) from the non-equispaced frequency samples. A great number of reconstruction methods exist for non-Cartesian image reconstruction. These methods are usually categorized into three families: (1) density compensation and gridding, (2) least square regression in $k$-space, and (3) iterative NUFFT. 
 	\subsubsection{Density compensation and gridding}
	The sampling density compensation method generated a tapering function $w$ (sampling density compensation function), which can be represented as the following stable iterations \cite{pipe1999sampling}:
	\begin{eqnarray}
	w_{i+1} = \frac{w_i}{{\bf AA^H}w_i}	
	\label{eq:pipe_density}	
	\end{eqnarray}
	Once $w$ is prepared, the sampling density compensation will be calculated by:
	\begin{eqnarray}
	x = {\bf A^H}(y\cdot w)
	\label{eq:sampling_density_compensation}
	\end{eqnarray}
	\subsubsection{Least square regression}
	Least square regression is a solution to the inverse problem of image reconstruction. Consider the following problem:
	\begin{eqnarray}     
     x = argmin_x\left\Vert y-{\bf A}x\right\Vert_2   
     \label{eq:basic_least_square}  
     \end{eqnarray}
     
	\begin{itemize}
    \item {\bf Conjugate gradient method (cg)}: A solution to Equation (\ref{eq:basic_least_square}) is the two-step solution:

     \begin{eqnarray}
     \label{eq:solver}
     k &=& min_k\left\Vert y -{\bf V}k\right\Vert_2 
     \end{eqnarray}
     The most expensive step is to find $k$ in Equation (\ref{eq:solver}). Once $k$ has been solved, the inverse of ${\bf FS}$ can be computed by the inverse FFT and then divided by the scaling factor:
     \begin{eqnarray}
	 x &=& S^{-1}F^{-1}k
     \end{eqnarray}

The conjugate gradient method is an iterative solution when the sparse matrix is a symmetric Hermitian matrix:
     \begin{eqnarray}
      {\bf V^H}y & = & {\bf V^HV}k,\ solve\ k \label{eq:solver_cg}     
     \end{eqnarray}
     
     Then each iteration generates the residue, which is used to compute the next value. The conjugate gradient method is also provided for heterogeneous systems. 
	\item {\bf Other iterative least square solvers}: 
Scipy \cite{Scipy} provides a variety of least square solvers, including lsmr, lsqr, bicg, bicgstab, gmres, lgmres. These solvers were integrated into the CPU version of PyNUFFT.
	\end{itemize}
		\subsubsection{Iterative NUFFT}
	 Iterative NUFFT reconstruction solves the inverse problem with various forms of image regularization. Due to the large size of the interpolation and FFT, iterative NUFFT is computationally expensive. 
	 Here, PyNUFFT is also optimized for iterative NUFFT reconstructions on heterogeneous systems. 
\begin{itemize}
 \item{\bf Pre-indexing for fast image gradient}:
 Total-variation is a basic image regularization which has been extensively used in image denosing and image reconstruction. Image gradient computes the difference between adjacent pixels, which is represented as follows: 
 \begin{eqnarray}
 \nabla_i{x} = x(...,a_i+1,...) - x(...,a_i,...)
 \end{eqnarray}
 where $a_i$ is the indexes of the $i$-th axis. Computing the image gradient requires image rolling, followed by subtracting the two images. However, multi-dimensional image rolling on heterogeneous is expensive and PyNUFFT adopts the pre-indexing to save the run-time. This pre-indexing procedure generates the indexes for the rolled image, and the indexes are offloaded to heterogeneous platforms before iterative algorithms begin. Thus, computing the image rolling is not needed during the iterations. 
 The advantage of this pre-indexing procedure is demonstrated in Figure \ref{fig:preindexing}, in which preindexing makes image gradient run faster on CPU and GPU. 
 \begin{figure}[!h]
\centering
\includegraphics[width=1.1\linewidth]{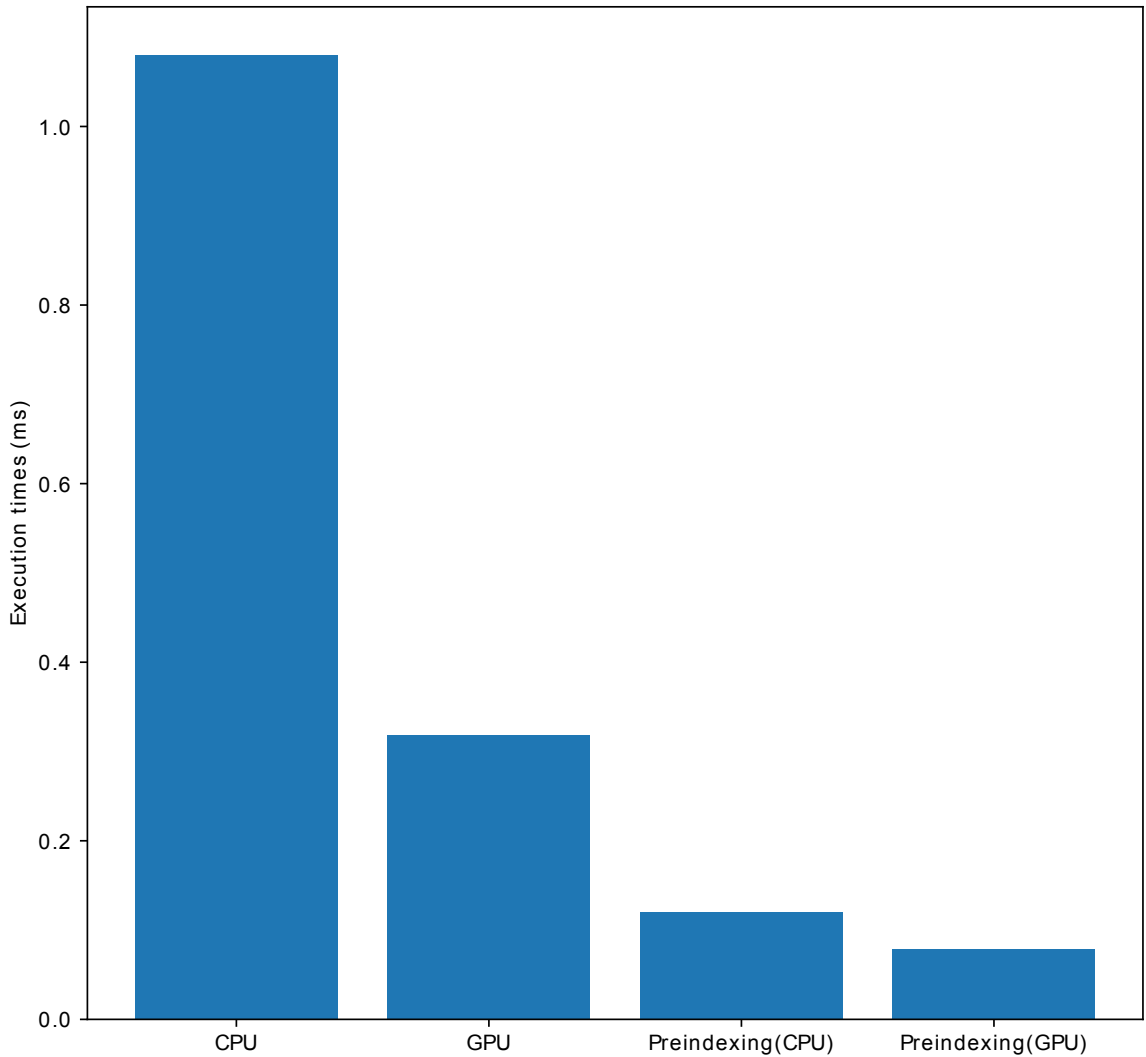} \caption{Preindexing for fast image gradient ($\nabla$ and $\nabla^T$). Image gradient can be slow if the indexes are to be calculated with each iteration. Preindexing can save the time needed for recalculation and image gradient during iterations.}
\label{fig:preindexing}
\end{figure}

 \item {\bf $\ell$1 total-variation regularized ordinary least square ({L1TV-OLS})}: The $\ell$1 total-variation regularized reconstruction includes the piece-wise smoothing into the reconstruction model.
	\begin{eqnarray}
	x=argmin_x	(\mu\left\Vert y - Ax \right\Vert_2 + \lambda TV(x) )
	\end{eqnarray} where $TV(x)$ is the total-variation of the image:
	\begin{eqnarray}
	TV(x) = \Sigma |\nabla_x x| +|\nabla_y x|
	\label{eq:TV}
	\end{eqnarray}
	Here, the $\nabla_x$ and $\nabla_y$ are directional gradient operators applied to the image domain along the $x$ and $y$ axes. 
	Equation (\ref{eq:TV}) is solved by the variable-splitting method and has previously been developed\cite{lin2015ismrmpico, lin2016ismrm, lin2017ismrm_pico}:

\begin{eqnarray}
K & = & \mu {\bf A^H A} + \lambda \nabla_x^T\nabla_x + \lambda \nabla_y^T\nabla_y
\end{eqnarray}
The inner iterations are as follows:
\begin{eqnarray}
rhs^k  & = &  \mu ({\bf A^H}y^k) + \lambda \nabla_x^T (d_x^k - b_x^k) + \lambda \nabla_y^T (d_y^k - b_y^k) \\
x^{k+1}  & = & K^{-1} rhs^{k} \label{eq:tvbreg_start}\\
d_x^{k+1} & = & shrink(\nabla_x x^{k+1} + b_x^k, 1/\lambda) \\
d_y^{k+1} & = & shrink(\nabla_y x^{k+1} + b_y^k, 1/\lambda) \\
b_x^{k+1} & = & b_x^k + (\nabla_x x^{k+1} - d_x^{k+1}) \\
b_y^{k+1} & = & b_y^k + (\nabla_y x^{k+1} - d_y^{k+1})
\end{eqnarray}

The outer iteration is:
\begin{eqnarray}
{\bf A^H}y^{k+1} = {\bf A^H}y^k + {\bf A^H}y -  {\bf A^HA}x^{k+1}
\label{eq:tvbreg_end}
\end{eqnarray}

	\item{\bf $\ell$1 total-variation regularized least absolute deviation (L1TV-LAD)}: Least absolute deviation (LAD) is a statistical regression model which is robust to non-stationary noise distribution \cite{Wang2013135}. It is possible to solve the  $\ell$1 total-variation regularized problem with the LAD cost function \cite{lin2017lad}: 
\begin{eqnarray}
	x=argmin_{x}(\mu\left\Vert y - Ax \right\Vert_1 + \lambda TV(x) )
	\end{eqnarray} where $TV(x)$ is the total-variation of the image. Note that LAD is the $\ell$1 norm of the data consistency. 

Here, a variable-splitting method is described. 
\begin{eqnarray}
K & = & \mu {\bf A^H A} + \lambda \nabla_x^T\nabla_x + \lambda \nabla_y^T\nabla_y
\end{eqnarray}
The inner iterations of $\ell$1-LAD are as follows:
\begin{eqnarray}
rhs^k  & = &  \mu ({\bf A^H}y^k + d_f^k -b_f^k) + \lambda \nabla_x^T (d_x^k - b_x^k) + \lambda \nabla_y^T (d_y^k - b_y^k) \\
x^{k+1}  & = & K^{-1} rhs^{k} \label{eq:lad_start}\\
d_x^{k+1} & = & shrink(\nabla_x x^{k+1} + b_x^k, 1/\lambda) \\
d_y^{k+1} & = & shrink(\nabla_y x^{k+1} + b_y^k, 1/\lambda) \\
d_f^{k+1} & = & shrink({\bf A^HA}x^{k+1} - {\bf A^H}y + b_y^k, 1/\mu) \label{eq:shrink_cost}\\
b_x^{k+1} & = & b_x^k + (\nabla_x x^{k+1} - d_x^{k+1}) \\
b_y^{k+1} & = & b_y^k + (\nabla_y x^{k+1} - d_y^{k+1}) \\
b_f^{k+1} & = & b_f^k + ({\bf A^HA}x^{k+1} - {\bf A^H}y - d_f^{k+1}) 
\end{eqnarray}

The outer iteration is:
\begin{eqnarray}
{\bf A^H}y^{k+1} = {\bf A^H}y^k + {\bf A^H}y -  {\bf A^HA}x^{k+1}
\label{eq:lad_end}
\end{eqnarray}
Note that Equation (\ref{eq:shrink_cost}) is a shrinkage function, which is quickly solved on CPU as well as on heterogeneous systems.

 \item{\bf Multi-coil regularized image reconstruction}: In multi-coil regularized image reconstruction, the selfadjoint NUFFT in Equation (\ref{eq:normalequation}) is extended to multi-channel data:
\begin{eqnarray}
{\bf A_{multi}^HA_{multi}}x & = & \sum_{i}^{Nc}{\bf A_{i}^HA_{i}}x \\
&=& \sum_{i}^{Nc}conj(c_i)\cdot{\bf A^HA}\cdot c_i\cdot x
\end{eqnarray} 
where coil-sensitivities ($c_i$) of multiple channels multiply each channel before the NUFFT ($\bf A$) and after the adjoint NUFFT ($\bf A^H$).
  
\end{itemize}

\subsection{Applications to brain MRI}

A 3T brain MRI template \cite{lalys2010construction} was used to simulate the non-Cartesian MRI acquisition. The image was resized to 512$\times$512 and the non-Cartesian $k$-space \cite{lin2015iterative} was used to simulate the data. 

\subsection{Benchmarks}

PyNUFFT was tested on a multi-core CPU cluster and a GPU instance of the cloud-based Linux web services. All the computations were completed with single float (FP32) complex numbers. The configuration of CPU and GPU systems were as follows. 

\begin{itemize}

\item {\bf Multi-core CPU}: The CPU instance (m4.16xlarge, Amazon Web Services) is equipped with 64 vCPUs (Intel E5 2686 v4) and 61 GB of memory. The number of vCPUs can be dynamically controlled by the CPU hotplug functionality of the Linux system, and computations are offloaded to the Intel OpenCL CPU device with 1 to 64 threads.

PyNUFFT was executed on the multi-core CPU instance with 1 - 64 threads. The PyNUFFT transforms were offloaded to the OpenCL CPU device and were executed 20 times. The times required for the transforms are compared with the run times on the single thread CPU.

Iterative reconstructions were also tested on the multi-core CPU. We measured the execution times of the conjugate gradient method, $\ell$1 total-variation regularized reconstruction and the $\ell$1 total-variation regularized LAD on the multi-core system.

\item  {\bf GPU}: The GPU instance (p2.xlarge, Amazon Web Services) is equipped with 4 vCPUs (Intel E5 2686 v4) and one Tesla K80 (NVIDIA, Santa Clara, CA, USA) with two GK210 GPUs. Each GPU is composed of 2496 parallel processing cores and 12 GB of memory. Computations are compiled and offloaded to one GPU by CUDA and OpenCL APIs. 

Transformations were repeated 20 times to measure the averaged run times. Iterative reconstructions were also tested on the K80 GPU, and the execution times of conjugate gradient method, $\ell$1 total-variation regularized reconstruction and $\ell$1 total-variation regularized LAD on the multi-core system were compared with iterative solvers on the CPU with one thread.

\end{itemize}

\section{Results}

\subsection{Applications to brain MRI}

\begin{figure}[!h]
\centering
\includegraphics[width=1.1\linewidth]{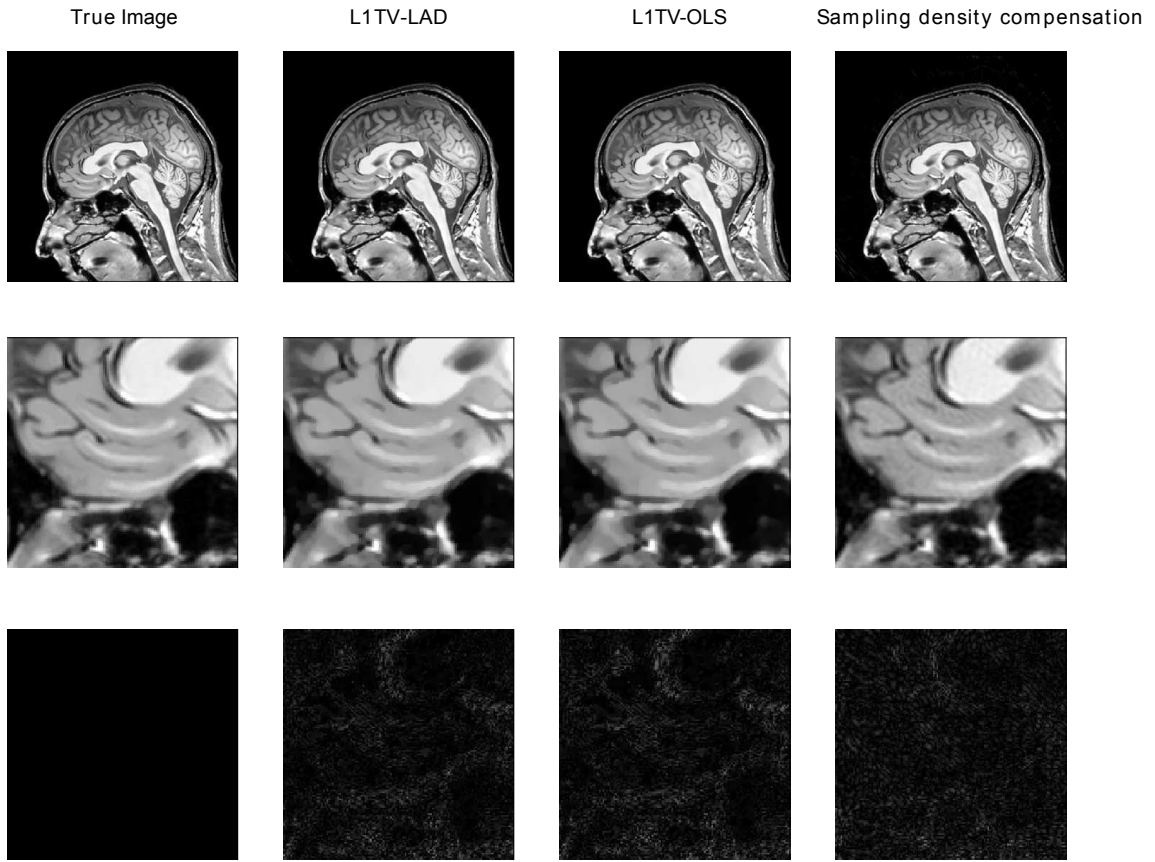} \caption{Simulated results of brain MRI. The brain template was reconstructed with two iterative NUFFT algorithms (L1TV-OLS and L1TV-LAD) and the sampling density compensation method. Compared with sampling density compensation method, fewer ripples can be seen in L1TV-OLS and L1TV-LAD.}
\label{fig:brain1}
 
\end{figure}

This work focuses on accelerating the PyNUFFT package on heterogeneous platforms, and the complete image quality assessment was not carried out in this paper. Some visual results of MRI reconstructions can be seen in Figure \ref{fig:brain1}. Iterative NUFFT using L1TV-OLS and L1TV-LAD result in fewer ripples in the brain structure than the sampling density compensation method. 

\subsection{Benchmarks}

\begin{itemize}

\begin{table}\centering
\caption{\label{table:accelerated_transforms}Run time (and acceleration) of subroutines and solvers}
\begin{tabular}{  c c c c c } 
 \hline
Operations & 1 vCPU  & 32 vCPUs & PyOpenCL K80 & PyCUDA K80  \\
 \hline \hline
 Scaling(${\bf S}$) & 709 $\mu$s & 557 $\mu$s (1.3$\times$) & 133 $\mu$s  (5.3$\times$) & 300 $\mu$s (2.4$\times$) \\ 
 FFT(${\bf  F}$) & 12.6ms & 2.03 ms (6.2$\times$) & 0.78 ms (16.2$\times$) & 1.12 ms (11.3$\times$)  \\ 
 Interpolation(${\bf V}$) & 25 ms & 2 ms (12$\times$) & 4 ms(6$\times$) & 4 ms (6$\times$) \\ 
 Gridding(${\bf V^H}$) & 24 ms & 2 ms (12$\times$) & 5 ms (4.8$\times$) & 4.5 ms (5.4$\times$)\\
 IFFT(${\bf F^H}$) & 16 ms & 3.1 ms (5.16$\times$) & 4.0 ms (4$\times$) & 3.5 (4.6$\times$) \\ 
 Rescaling (${\bf S^H}$) & 566 $\mu$s & 595 $\mu$s (0.95$\times$)& 122 $\mu$s (4.64$\times$) & 283 $\mu$s (2$\times$) \\
 ${\bf V^HV}$ & 4.79 ms & 2.62 ms (1.83$\times$) & 0.18 ms (26$\times$) &0.15 ms (31$\times$) \\ 
  \hline \hline
 Forward (${\bf A}$) & 39 ms & 5 ms (7.8$\times$)& 6 ms (6.5$\times$) & 6 ms (6.5$\times$) \\
 Adjoint (${\bf A^H}$) & 38 ms & 6 ms (6.3$\times$) & 7 ms (5.4$\times$) &6 ms (6.3$\times$) \\ 
 Selfadjoint (${\bf A^HA}$) & 34.7 ms & 11 ms (3.15$\times$) & 9 ms (3.86$\times$) &8 ms (4.34$\times$) \\     
 \hline \hline
 Solvers & 1 vCPU  & 32 vCPUs & PyOpenCL K80 & PyCUDA K80  \\
\hline
Conjugate gradient & 11.8 s & 2.97 s (4$\times$) & 1.32 s (8.9$\times$) & 1.89 s (6.3$\times$) \\  
L1TV-OLS &14.7 s & 3.26 s (4.5$\times$) & 1.79 s (8.2$\times$) & 1.68 s (8.7$\times$)
\\
L1TV-LAD &15.1 s & 3.62 s (4.2$\times$) & 1.93 s (7.8$\times$) & 1.78 s (8.5$\times$) \\
 \hline
\end{tabular}
\end{table}


\begin{figure}[b]
\centering
\includegraphics[width=1.1\linewidth]{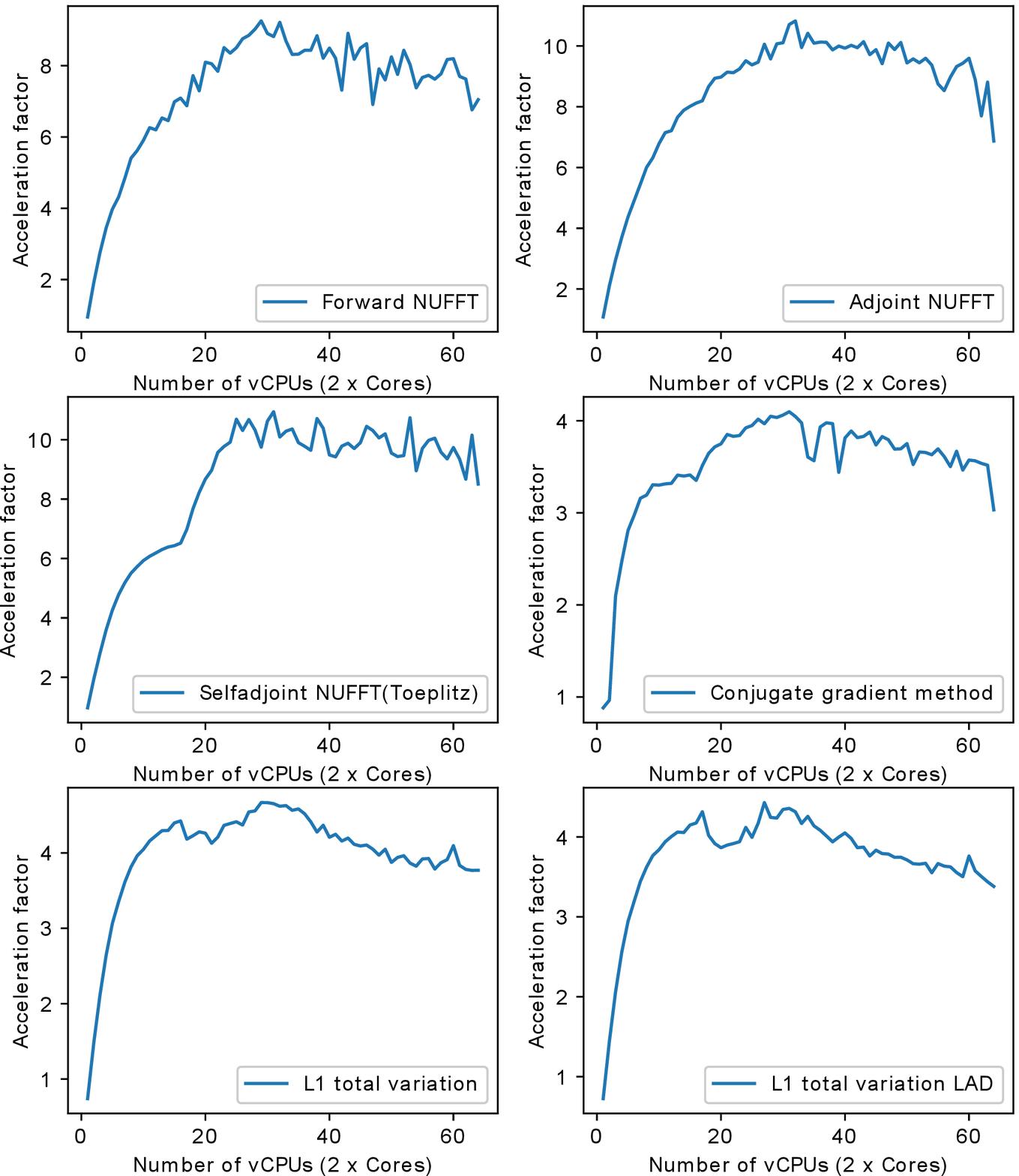} \caption{Acceleration factors against the number of threads (vCPUs for Amazon Web Services). The optimal performance occurs around 30 - 32 threads. There is no substantial benefits for more than 32 threads.}
\label{fig:acceleration_on_64cores}
\end{figure}

\item {\bf Multi-core CPU}: 
Table \ref{table:accelerated_transforms} listed the detailed run-times of the different stages of forward NUFFT, adjoint NUFFT, and selfadjoint NUFFT (Toeplitz).
The overall execution speed of PyNUFFT was faster on the multi-core CPU platform than single thread CPU. Yet the acceleration factors of each stage varied from 0.95 (no acceleration) to 15.6.  Compared with computations on single thread, 32 threads accelerate interpolation and gridding by a factor of 12, and the FFT and IFFT were accelerated by a factor of 6.2 - 15.6. 

The benefits of 32 threads are limited for certain computations, including scaling, rescaling and interpolation-gridding ($\bf V_HV$)). In these computations, the acceleration factors of 32 threads range from 0.95 to 1.85. This limited performance gain is due to the high efficiency of single thread CPU, which is already fast enough and leaves limited room for improvement. In particular, the integrated interpolation-gridding ($\bf V_HV$) is already 10 times faster than the separate interpolation and regridding sequence. On a single-thread CPU, $\bf V_HV$ requires only 4.79 ms, whereas the separate interpolation ($\bf V$) and gridding ($\bf V^H$) require 49 ms. In this case, 32 threads deliver only an extra 83\% of performance to $\bf V_HV$.

Figure \ref{fig:acceleration_on_64cores} illustrates the acceleration on the multi-core CPU against the single thread CPU. The performance of PyNUFFT improved by a factor of 5 - 10 when the number of threads increases from 1 to 20, and the software achieves peak performance with 30 - 32 threads (equivalent to 15 - 16 physical CPU cores). More than 32 threads bring no substantial benefits in performance.

Forward NUFFT, adjoint NUFFT and selfadjoint NUFFT (Toeplitz) are accelerated on 32 threads with an acceleration factor of 7.8 - 9.5. The acceleration factors of iterative solvers (conjugate gradient method, $\ell$1 total-variation regularized least square (L1TV-OLS) and $\ell$1 total-variation regularized least absolute deviation (L1TV-LAD)) are from 4.2 - 5.

\item {\bf GPU}: 

Table \ref{table:accelerated_transforms} shows that GPU delivers a generally faster PyNUFFT transform, with the acceleration factors ranging from 2 to 31. 

Scaling and rescaling have led to a moderate degree of acceleration. The most significant acceleration occurs in the interpolation-gridding ($\bf V_HV$) in which GPU is 26 - 31 times faster than single-thread CPU. This significant acceleration is faster than the acceleration factors for separate interpolation ($\bf V$, with 6$\times$ acceleration) and gridding ($\bf V^H$ with 4 - 4.6$\times$ acceleration).

Forward NUFFT, adjoint NUFFT and selfadjoint NUFFT (Toeplitz) are accelerated on K80 GPU by 5.4 - 13. Iterative solvers on GPU are 6.3 - 8.9 faster than single-thread, and about 2$\times$ faster than 32 threads.

\end{itemize}

\section{Discussion\label{sec:discussion}}

\subsection{Related work}  
  
Different kernel functions (interpolators) are available in previous NUFFT implementations, including: (1)the min-max interpoaltor \cite{fessler2003nonuniform}, (2) the fast radial basis functions \cite{Potts2004nffts,keiner2009using}, (3)least square interpolator \cite{song2009least}, (4)least mean-square error interpolato \cite{yang2014mean}, (5) fast Gaussian summation \cite{greengard2004accelerating}, (6)  Kaiser-Bessel function \cite{gpuNUFFT2014}, (7) linear system transfer function or inverse reconstruction \cite{liu2005fast, uecker2010nonlinear}.

The NUFFTs have been implemented in different programming languages: (1) Matlab (Mathworks Inc, MA, USA)\cite{fessler2003nonuniform, yang2014mean, gpuNUFFT2014, mathhew2009nufftmatlab}; (2) C++ \cite{agile2012}; (3) CUDA \cite{gpuNUFFT2014, agile2012}; (4) Fortran \cite{greengard2004accelerating}; (5) OpenACC using PGI compiler (PGI Compilers \& Tools) \cite{powergrid2016}.

NUFFT has been accelerated on single and multiple GPUs. Fast iterative NUFFT using the Kaiser-Bessel function was accelerated on GPU with total-variation regularization \cite{agile2012} and generalized total-variation regularization \cite{gpuNUFFT2014}. A real-time inverse reconstruction is developed in Sebastinan et al. \cite{Sebastian2017GPU} and Murphy et al. \cite{murphy2012fast}, but this inverse reconstruction does not perform the full interpolation and gridding during iterations. The patent of Nadar et al.\cite{nadar2013multi} describes a custom multi-GPU buffer to improve the memory access for image reconstruction with non-uniform $k$-space. An mripy package applies the Numba compiler in its NUFFT with Gaussian kernel (https://github.com/peng-cao/mripy). 

Other than NUFFT, iterative discrete Fourier transform (DFT) is slower than NUFFT, and GPUs can bring enormous acceleration factors to iterative reconstruction \cite{Stone:2008:AAM:1366230.1366276, gai2013more}. Still, iterative DFTs on GPUs may be slower than iterative NUFFTs on GPUs. 

\subsection{Discussions of PyNUFFT}  
 
The NUFFT transforms (forward, adjoint, and Toeplitz) have been accelerated on multi-core CPU and GPU. In particular, the benefits of fast iterative solvers (including least square and iterative NUFFT) have been shown in the results of benchmarks. The image reconstruction times (with 100 iterations) for one 256$\times$256 image are less than 4 seconds on a 32 thread CPU platform and less than 2 seconds on the GPU platform.

The current PyNUFFT has been tested with computations using the single-precision floating numbers (FP32). However, the number of FP64 units on GPU is only a fraction of the number of FP32 units, which will reduce the performance with FP64 and slow down the performance of PyNUFFT (with 1/3 of the FP32 performance).



\subsection{Conclusion}

An open-source PyNUFFT package can accelerate non-Cartesian image reconstruction on multi-core CPU and GPU platforms. The acceleration factors are 6.3 - 9.5$\times$ on a 32 thread CPU platform and 5.4 - 13$\times$ on a Tesla K80 GPU. The iterative solvers with 100 iterations can be completed within 4 seconds on the 32 thread CPU platform and 2 seconds on the GPU.
\section{Acknowledgements}  
     The work was supported by the Ministry of Science and Technology (MOST) Taiwan, partly by the Cambridge Commonwealth, European and International Trust (Cambridge, UK), as well as the Ministry of Education, Taiwan. The benchmarks were carried out on Amazon Web Services provided by AWS Educate credit. J.-M. Lin declares no conflict of interest.
\section{References}  
\bibliographystyle{elsarticle-num}
\bibliography{whipple.bib} 

\begin{thebibliography}{10}
\expandafter\ifx\csname url\endcsname\relax
  \def\url#1{\texttt{#1}}\fi
\expandafter\ifx\csname urlprefix\endcsname\relax\def\urlprefix{URL }\fi
\expandafter\ifx\csname href\endcsname\relax
  \def\href#1#2{#2} \def\path#1{#1}\fi

\bibitem{klockner2012pycuda}
A.~Kl{\"o}ckner, N.~Pinto, Y.~Lee, B.~Catanzaro, P.~Ivanov, A.~Fasih, {PyCUDA
  and PyOpenCL}: A scripting-based approach to {GPU} run-time code generation,
  Parallel Computing 38~(3) (2012) 157--174.

\bibitem{lin2017bbmc_pynufft}
J.-M. Lin, H.-W. Chung, {PyNUFFT: python non-uniform fast Fourier transform for
  MRI}, in: Building Bridges in Medical Sciences 2017, St John's College,
  Cambridge CB2 1TP, UK, 2017.

\bibitem{reikna}
B.~Opanchuk, \href{{http://reikna.publicfields.net/}}{{Reikna, a pure Python
  {GPGPU} library}}, {http://reikna.publicfields.net/}, Accessed: 2017-10-09.
\newline\urlprefix\url{{http://reikna.publicfields.net/}}

\bibitem{gplv3}
{Free Software Foundation}, \href{http://www.gnu.org/licenses/gpl.html}{{GNU
  General Public License}}.
\newline\urlprefix\url{http://www.gnu.org/licenses/gpl.html}

\bibitem{fessler2003nonuniform}
J.~Fessler, B.~P. Sutton, Nonuniform fast {F}ourier transforms using min-max
  interpolation, IEEE Transactions on Signal Processing 51~(2) (2003) 560--574.

\bibitem{fessler2005toeplitz}
J.~Fessler, S.~Lee, V.~T. Olafsson, H.~R. Shi, D.~C. Noll, Toeplitz-based
  iterative image reconstruction for {MRI} with correction for magnetic field
  inhomogeneity, IEEE Transactions on Signal Processing 53~(9) (2005)
  3393--3402.

\bibitem{pipe1999sampling}
J.~G. Pipe, P.~Menon, Sampling density compensation in {MRI}: Rationale and an
  iterative numerical solution, Magnetic Resonance in Medicine 41~(1) (1999)
  179--186.

\bibitem{Scipy}
E.~Jones, T.~Oliphant, P.~Peterson, et~al.,
  \href{http://www.scipy.org/}{{SciPy}: Open source scientific tools for
  {Python}} (2001--).
\newline\urlprefix\url{http://www.scipy.org/}

\bibitem{lin2015ismrmpico}
J.-M. Lin, A.~J. Patterson, H.-C. Chang, T.-C. Chuang, H.-W. Chung, M.~J.
  Graves, Whitening of colored noise in {PROPELLER} using iterative regularized
  {PICO} reconstruction, in: Proceedings of the 23rd Annual Meeting of
  International Society for Magnetic Resonance in Medicine, Toronto, Canada,
  2015, p. 3738.

\bibitem{lin2016ismrm}
J.-M. Lin, A.~J. Patterson, C.-W. Lee, Y.-F. Chen, T.~Das, D.~Scoffings, H.-W.
  Chung, J.~Gillard, M.~Graves, Improved identification and clinical utility of
  pseudo-inverse with constraints {(PICO)} reconstruction for {PROPELLER}
  {MRI}, in: Proceedings of the 24th Annual Meeting of International Society
  for Magnetic Resonance in Medicine, Singapore, 2016, p. 1773.

\bibitem{lin2017ismrm_pico}
J.-M. Lin, S.-Y. Tsai, H.-C. Chang, H.-W. Chung, H.~C. Chen, Y.-H. Lin, C.-W.
  Lee, Y.-F. Chen, D.~Scoffings, T.~Das, J.~H. Gillard, A.~J. Patterson, M.~J.
  Graves, {Pseudo-inverse constrained (PICO) reconstruction reduces colored
  noise of PROPELLER and improves the gray-white matter differentiation}, in:
  Proceedings of the 25th Annual Meeting of International Society for Magnetic
  Resonance in Medicine, Honolulu, HI, USA, 2017, p. 1524.

\bibitem{Wang2013135}
L.~Wang, The penalized {LAD} estimator for high dimensional linear regression,
  Journal of Multivariate Analysis 120 (2013) 135 -- 151.
\newblock \href {http://dx.doi.org/http://doi.org/10.1016/j.jmva.2013.04.001}
  {\path{doi:http://doi.org/10.1016/j.jmva.2013.04.001}}.

\bibitem{lin2017lad}
J.-M. Lin, H.-C. Chang, T.-C. Chao, S.-Y. Tsai, A.~Patterson, H.-W. Chung,
  J.~Gillard, M.~Graves, {L1-LAD}: Iterative {MRI} reconstruction using {L1}
  constrained least absolute deviation, in: European Society for Magnetic
  Resonance in Medicine and Biology (ESMRMB) 2017 Congress, 2017.

\bibitem{lalys2010construction}
F.~Lalys, C.~Haegelen, J.-C. Ferre, O.~El-Ganaoui, P.~Jannin, Construction and
  assessment of a 3-{T} {MRI} brain template, Neuroimage 49~(1) (2010)
  345--354.

\bibitem{lin2015iterative}
J.-M. Lin, A.~J. Patterson, H.-C. Chang, J.~H. Gillard, M.~J. Graves, An
  iterative reduced field-of-view reconstruction for periodically rotated
  overlapping parallel lines with enhanced reconstruction {PROPELLER} {MRI},
  Medical Physics 42~(10) (2015) 5757--5767.

\bibitem{Potts2004nffts}
D.~Potts, G.~Steidl, Fast summation at nonequispaced knots by {NFFT}s., SIAM
  Journal on Science Computing 24 (2004) 2013--2037.

\bibitem{keiner2009using}
J.~Keiner, S.~Kunis, D.~Potts, Using {NFFT} 3 --- a software library for
  various nonequispaced fast {F}ourier transforms, ACM Transactions on
  Mathematical Software (TOMS) 36~(4) (2009) 19.

\bibitem{song2009least}
J.~Song, Y.~Liu, S.~L. Gewalt, G.~Cofer, G.~A. Johnson, Q.~H. Liu,
  {Least-square NUFFT methods applied to 2-D and 3-D radially encoded MR image
  reconstruction}, IEEE Transactions on Biomedical Engineering 56~(4) (2009)
  1134--1142.

\bibitem{yang2014mean}
Z.~Yang, M.~Jacob, Mean square optimal {NUFFT} approximation for efficient
  non-{C}artesian {MRI} reconstruction, Journal of Magnetic Resonance 242
  (2014) 126--135.

\bibitem{greengard2004accelerating}
L.~Greengard, J.-Y. Lee, Accelerating the nonuniform fast {F}ourier transform,
  SIAM review 46~(3) (2004) 443--454.

\bibitem{gpuNUFFT2014}
F.~Knoll, A.~Schwarzl, C.~S.~D. Diwoky, {gpuNUFFT – An open-source GPU
  library for 3D gridding with direct Matlab Interface.}, in: Proc ISMRM, 2014,
  p. 4297.

\bibitem{liu2005fast}
C.~Liu, M.~Moseley, R.~Bammer, Fast {SENSE} reconstruction using linear system
  transfer function., in: Proceedings of the International Society of Magnetic
  Resonance in Medicine, 2005, p. 689.

\bibitem{uecker2010nonlinear}
M.~Uecker, S.~Zhang, J.~Frahm, Nonlinear inverse reconstruction for real-time
  {MRI} of the human heart using undersampled radial {FLASH}, Magnetic
  Resonance in Medicine 63~(6) (2010) 1456--1462.

\bibitem{mathhew2009nufftmatlab}
M.~Ferrara, Implements 1{D}-3{D} {NUFFT}s via fast {G}aussian gridding., matlab
  Central,
  http://www.mathworks.com/matlabcentral/fileexchange/25135-nufft--nfft--usfft
  (2009).

\bibitem{agile2012}
K.~Bredies, F.~Knoll, M.~Freiberger, H.~Scharfetter, R.~Stollberger, The agile
  library for biomedical image reconstruction using {GPU} acceleration,
  Computing in Science and Engineering 15 (2013) 34--44.

\bibitem{powergrid2016}
A.~Cerjanic, J.~L. Holtrop, G.~C. Ngo, B.~Leback, G.~Arnold, M.~V. Moer,
  G.~LaBelle, J.~A. Fessler, B.~P. Sutton, {PowerGrid: A open source library
  for accelerated iterative magnetic resonance image reconstruction}, in: Proc
  ISMRM, 2016, p. 525.

\bibitem{Sebastian2017GPU}
S.~Schaetz, D.~Voit, J.~Frahm, M.~Uecker,
  {Accelerated computing in magnetic
  resonance imaging: Real-time imaging using non-linear inverse
  reconstruction}, CoRR.
\newline\urlprefix\url{http://arxiv.org/abs/1701.08361}

\bibitem{murphy2012fast}
M.~Murphy, M.~Alley, J.~Demmel, K.~Keutzer, S.~Vasanawala, M.~Lustig,
  Fast-{SPIRiT} compressed sensing parallel imaging {MRI}: Scalable parallel
  implementation and clinically feasible runtime, IEEE Transactions on Medical
  Imaging 31~(6) (2012) 1250--1262.

\bibitem{nadar2013multi}
M.~S. Nadar, S.~Martin, A.~Lefebvre, J.~Liu, Multi-{GPU} {FISTA} implementation
  for {MR} reconstruction with non-uniform k-space sampling, uS Patent App.
  14/031,374 (2013).

\bibitem{Stone:2008:AAM:1366230.1366276}
S.~S. Stone, J.~P. Haldar, S.~C. Tsao, W.-m.~W. Hwu, Z.-P. Liang, B.~P. Sutton,
  \href{http://doi.acm.org/10.1145/1366230.1366276}{Accelerating advanced {MRI}
  reconstructions on {GPU}s}, in: Proceedings of the 5th Conference on
  Computing Frontiers, CF '08, ACM, New York, NY, USA, 2008, pp. 261--272.
\newblock \href {http://dx.doi.org/10.1145/1366230.1366276}
  {\path{doi:10.1145/1366230.1366276}}.
\newline\urlprefix\url{http://doi.acm.org/10.1145/1366230.1366276}

\bibitem{gai2013more}
J.~Gai, N.~Obeid, J.~L. Holtrop, X.-L. Wu, F.~Lam, M.~Fu, J.~P. Haldar, W.~H.
  Wen-mei, Z.-P. Liang, B.~P. Sutton, More {IMPATIENT}: A gridding-accelerated
  {T}oeplitz-based strategy for non-{C}artesian high-resolution 3{D} {MRI} on
  {GPU}s, Journal of Parallel and Distributed Computing 73~(5) (2013) 686--697.

\end{thebibliography}

\end{document}